\documentclass[aps,prb,10pt,twocolumn,longbibliography,superscriptaddress]{revtex4-2}
\usepackage{multirow}
\usepackage{graphicx}
\usepackage{dcolumn}
\usepackage{bm}
\usepackage{hyperref}
\hypersetup{
    colorlinks=true,
    urlcolor= blue,
    citecolor=blue,
linkcolor= blue}
\usepackage{amsmath,amssymb,amsfonts}%
\usepackage{amsthm}%
\usepackage{mathrsfs}%
\usepackage{appendix}

\begin{document}

\preprint{APS/123-QED}

\title{Magnetic-field-free nonreciprocal transport in
graphene multi-terminal Josephson junctions}

\author{Fan Zhang}\thanks{These authors contributed equally to this work}
\affiliation{Department of Physics, The Pennsylvania State University, University Park, PA 16802, USA}

\author{Asmaul Smitha Rashid}\thanks{These authors contributed equally to this work}
\affiliation{Department of Electrical Engineering, The Pennsylvania State University, University Park, PA 16802, USA}

\author{Mostafa Tanhayi Ahari}
\affiliation{Materials Research Laboratory, The Grainger College of Engineering, University of Illinois, Urbana-Champaign, IL 
61801, USA}

\author{George J. de Coster}
\affiliation{DEVCOM Army Research Laboratory, 2800 Powder Mill Rd, Adelphi, MD, 20783, USA}

\author{Takashi Taniguchi}
\affiliation{International Center for Materials, Nanoarchitectonics, National
Institute for Materials Science, 1-1 Namiki, Tsukuba 305-0044, Japan.}

\author{Kenji Watanabe}
\affiliation{Research Center for Functional Materials, Institute for Materials
Science, 1-1 Namiki, Tsukuba 305-0044, Japan}

\author{Matthew J. Gilbert}
\affiliation{Department of Electrical Engineering, University of Illinois, Urbana-Champaign, IL 61801, USA}
\affiliation{Materials Research Laboratory, The Grainger College of Engineering, University of Illinois, Urbana-Champaign, IL 61801, USA}

\author{Nitin Samarth}\thanks{Corresponding authors: nsamarth@psu.edu, mzk463@psu.edu}
\affiliation{Department of Physics, The Pennsylvania State University, University Park, PA 16802, USA}
\affiliation{Department of Materials Science and Engineering, The Pennsylvania
State University, University Park, PA 16802, USA}

\author{Morteza Kayyalha}\thanks{Corresponding authors: nsamarth@psu.edu, mzk463@psu.edu}
\affiliation{Department of Electrical Engineering, The Pennsylvania State University, University Park, PA 16802, USA}

\date{\today}

\begin{abstract}
Nonreciprocal superconducting devices have attracted growing interest in recent years as they potentially enable 
directional charge transport for applications in superconducting quantum circuits.  Specifically, the superconducting diode effect has been explored in two-terminal devices that exhibit superconducting transport in one current direction while showing dissipative transport in the opposite direction. 
Here, we exploit multi-terminal Josephson junctions (MTJJs) to engineer magnetic-field-free nonreciprocity in multi-port networks. We show that when treated as a two-port electrical network, a three-terminal Josephson junction (JJ) with an asymmetric graphene region exhibits reconfigurable two-port nonreciprocity. We observe nonreciprocal (reciprocal) transport between superconducting terminals with broken (preserved) spatial mirror symmetry. We explain our observations by considering a circuit-network of JJs with different critical currents.
\end{abstract}

\maketitle

\section{Introduction}
In two-terminal electronic devices, the concept of reciprocity implies a symmetry relation between the applied current and 
measured voltage. In other words, the resistance remains the same if the polarity of the current source is reversed 
from positive to negative \cite{onsager1931reciprocal}. Violating this fundamental symmetry in semiconductor technology has led to the development of a wide range of established device technologies such as diodes, transistors, rectifiers, and photodetectors 
\cite{scaff1947development,shockley1949theory,monroy2003wide,sze2008semiconductor}. In superconducting devices, engineering nonreciprocity requires simultaneous breaking of time-reversal and inversion symmetries 
\cite{hu2007proposed,wakatsuki2018nonreciprocal,hoshino2018nonreciprocal,chen2018asymmetric,davydova2022universal,daido2022intrinsic,scammell2022theory,yuan2022supercurrent,ilic2022theory,he2022phenomenological,zhang2022general}. Current experimental demonstrations of nonreciprocity in superconductors have focused on two-terminal configurations where nonreciprocity emerges as the superconducting diode effect (SDE) \cite{ando2020observation,diez2021magnetic,shin2021magnetic,baumgartner2022supercurrent,narita2022field,jeon2022zero,bauriedl2022supercurrent,pal2022josephson,wu2022field,bauriedl2022supercurrent,lin2022zero,matsuo2023josephson}. In these two-terminal devices, the SDE has been identified by demonstrating asymmetric critical currents for the positive and negative current polarities, i.e., $I_c^+ \neq I_c^-$.

The SDE has also been reported in multi-terminal Josephson junctions (MTJJs) \cite{gupta2022superconducting,chiles2022non}. These studies, however, have treated MTJJs as two-terminal devices and reported nonreciprocity either under a small magnetic field \cite{gupta2022superconducting} or with an applied current bias \cite{chiles2022non}. Here, we construct devices based on MTJJs to demonstrate reconfigurable and magnetic-field-free nonreciprocal transport in multi-port networks. We find that in MTJJs, a current $I_j$  applied to terminal $j$  results in a critical current $I_{c,kj}$ at terminal $k$.  We observe two-port nonreciprocal transport (i.e., $I_{c,jk} \neq I_{c,kj}$), if the spatial mirror symmetry between terminals $j$ and $k$ is broken. Notably, we show that this two-port nonreciprocity is polarity independent ($I_{c,jk}^\pm \neq I_{c,kj}^\pm$) and emerges without any magnetic field or bias current. We design devices with asymmetric channel regions that allow us to break spatial mirror symmetry between specific terminals. We further control the efficiency of nonreciprocal transport by controlling the degree of asymmetry in the channel. Furthermore, using a circuit-network model of Josephson junctions (JJs), we show that the efficiency depends on the asymmetry of the JJs and reaches a maximum value of $30\%$. Our model is materials agnostic and applies to a broad range of junctions with sinusoidal CPR’s, regardless of the Fermi surface or the band structure of the normal material \cite{gupta2022superconducting}.

MTJJs have recently attracted increasing interest due to their potential to create novel superconducting properties such as quartet pairings 
\cite{Linder,pfeffer2014subgap,cohen2018nonlocal,draelos2019supercurrent,pankratova2020multiterminal,Graziano2020gate-tunable,Arnault2022dynamical,huang2022evidence,Grazino2022selective,melin2023quantum,melin2023proposal,Zhang2023Andreev} and to emulate higher dimensional topological phases 
\cite{riwar2016multi,strambini2016omega,meyer2017nontrivial}. Here, We treat a MTJJ with $n$ superconducting terminals as a ($n-$1)-port network. The reciprocity relation for multi-port networks states that a network is reciprocal when a voltage $V_k$ (at port $k$) produced by a current $I_j$ (injected to port $j$) remains the same if the current and voltage leads are exchanged \cite{onsager1931reciprocal,casimir1945onsager,philips1958method,buttiker1986four}. Figure~\ref{P1.png}A schematically demonstrates this reciprocity relation in a two-port network. 
Following this definition, we find that in MTJJs, a current $I_j$ applied to terminal $j$ results in a critical current ($I_{c,kj}$) at terminal $k$. We define this critical current as the maximum current $I_j$ for which $V_k = 0$. Likewise, a current $I_k$ applied to terminal $k$ results in a critical current $I_{c,jk}$ at terminal $j$. Therefore, we consider the MTJJ nonreciprocal if $I_{c,jk} \neq I_{c,kj}$. We explore this nonreciprocity in three- and four-terminal graphene JJs. We intentionally create a graphene channel with an asymmetric geometry to break the spatial mirror symmetry between some of the terminals. In these asymmetric MTJJs, we observe reconfigurable two-port nonreciprocity: transport is nonreciprocal between terminals with broken spatial mirror symmetry, whereas it is reciprocal between terminals with preserved spatial mirror symmetry. 
We use a circuit-network of JJs with different critical currents to explain the observed nonreciprocity in our MTJJs. 


\section{Device Configuration}

We fabricate our MTJJs on hBN/graphene/hBN van der Waals heterostructures which are edge-contacted by Ti (10 nm)/Al (100 nm) superconducting electrodes. Figure~\ref{P1.png}B shows an atomic force microscope (AFM) image of a representative three-terminal JJ. The graphene channel in this device is narrower ($\sim 400$ nm) between terminals (1, 3), whereas it is wider ($\sim 700$ nm) between terminals (1, 2) and (2, 3). The inset of Fig.~\ref{P1.png}B shows the circuit-network of JJs representing our three-terminal device. We use three JJs (JJ1, JJ2, JJ3) between each terminal pair with corresponding critical currents ($I_{c1}$, $I_{c2}$, $I_{c3}$). Because the graphene region has an asymmetric geometry, the critical current of JJ3 ($I_{c3}$) is lager than the critical current of JJ1 ($I_{c1}$) and JJ2 ($I_{c2}$), whereas JJ1 and JJ2 have approximately similar critical currents, i.e., $I_{c3} > I_{c1} \approx I_{c2}$. 
We characterize our three-terminal JJs using two different bias-current configurations. In the first configuration (Config.~1), we ground terminal 2 and apply currents $I_1$ and $I_3$ to terminals 1 and 3, respectively. In the second configuration (Config.~2), we ground terminal 3 and apply currents $I_1$ and $I_2$ to terminals 1 and 2, respectively. 
We perform all the measurements at gate voltage $V_g = 30$ V, magnetic field $B=0$ T, and temperature $T= 250$ mK, unless specified otherwise. 
We choose $T= 250$ mK to avoid Joule heating and stochastic switching noise \cite{fulton1974lifetime,clarke1988quantum} (see Appendix ~\ref{Stochastic switching} for more details). 
\section{Reconfigurable Multiport Nonreciprocity}
\begin{figure}
\centering
\includegraphics[width=9cm]{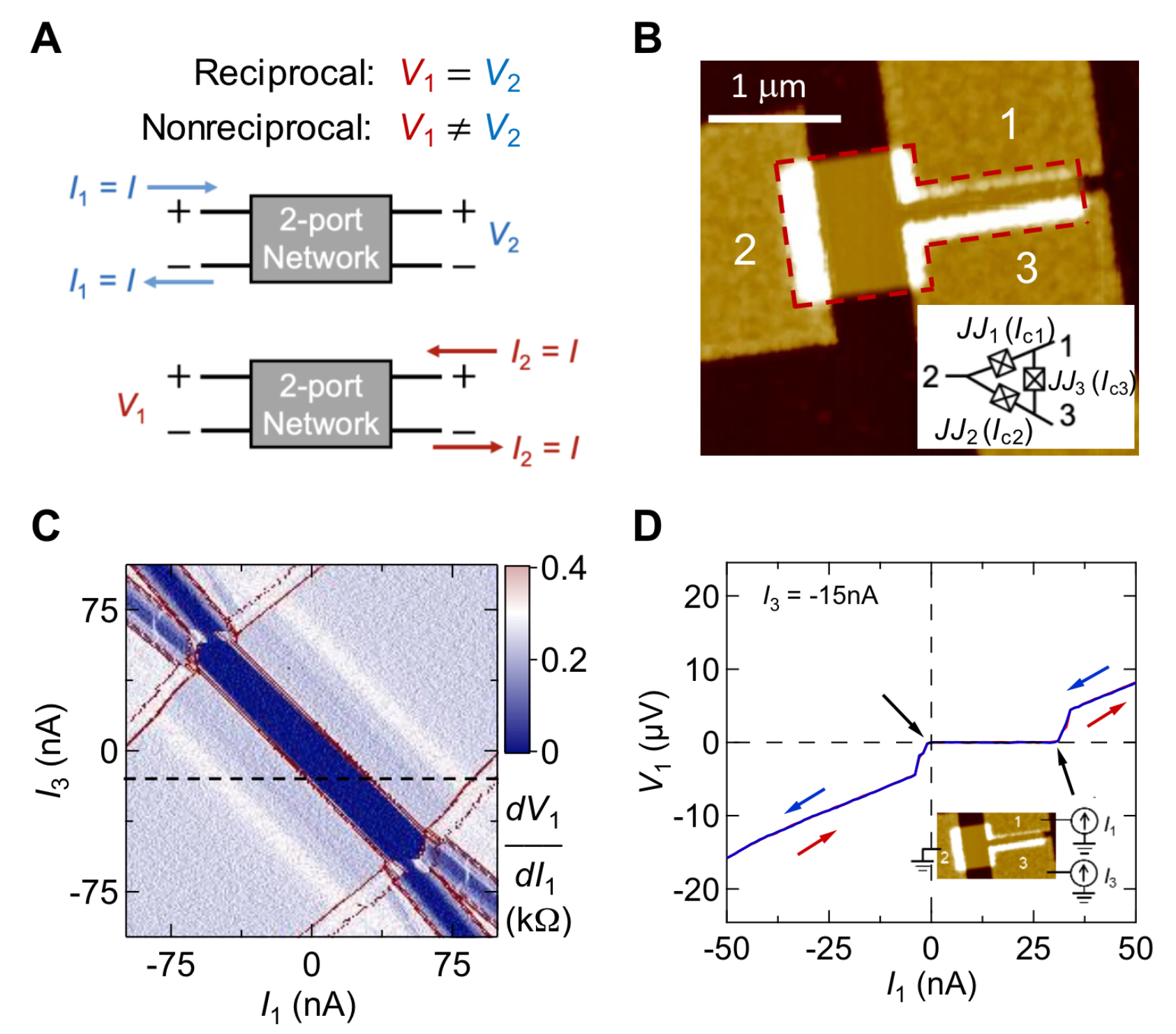
}
\caption{\label{P1.png} (\textbf{A}) 
A schematic of a two-port network. A current $I$ applied to port 1 ($I_1=I$) while  $I_2 = 0$ will produce a voltage $V_2$ at port 2. Likewise, the same current $I$ applied to port 2 ($I_2=I$) while  $I_1 = 0$ will produce a voltage $V_1$ at port 1. The network is reciprocal if $V_1 = V_2$, whereas it is nonreciprocal if $V_1 \neq V_2$.
(\textbf{B}) 
An AFM image of a representative three-terminal JJ. The JJ is made of hBN/graphene/hBN heterostructure (marked by the red dashed contour) edge contacted with Ti/Al superconducting terminals. 
The inset is a schematic of the circuit-network of coupled JJs utilized to model the three-terminal JJ. 
(\textbf{C}) 
Color map of the differential resistance $dV_{1}/dI_{1}$ versus $I_{1}$ and $I_{3}$. 
(\textbf{D}) 
A horizontal cut of $V_{1}$ vs $I_{1}$ along the black dashed line ($I_{3} = \textit{-}15$ nA) in panel (\textbf{C}). 
Red and blue arrows mark $I_1$ sweep directions. The black arrows mark the position of the critical current for positive and negative currents. All the measurements are performed at $V_g = 30$ V, $B = 0$ T, and $T = 250$ mK. 
}
\end{figure}

Figure~\ref{P1.png}C shows a color map of the differential resistance $dV_1/dI_1$ as a function of the bias currents $I_1$ and $I_3$ measured in the three-terminal JJ using Config. 1. The boundary of the dark blue region marks the critical current contour (CCC). We observe that the CCC in this configuration is asymmetric; it is titled by $\sim 45^{\circ}$ along the $V_1 - V_3 = 0$ direction. This is because the critical current of JJ3 ($I_{c3}$) is larger compared to the critical currents of JJ1 ($I_{c1}$) and JJ2 ($I_{c2}$); see Fig.~\ref{P1.png}B. We also observe that at a non-zero value of $I_3$, the critical current along the $I_1$ direction becomes asymmetric for the positive and negative current polarities. Here, we define the critical current as the maximum of the total current in terminal 1 that results in $V_1=0$.  
To highlight this asymmetry, Fig.~\ref{P1.png}D plots a cut of $V_1$-$I_1$ along the horizontal dashed line in Fig.~\ref{P1.png}C at $I_3 = -15$ nA. 
We observe that the critical current for the positive current polarity is around $31$ nA, whereas the critical current for the negative current polarity is around $0$ nA. We also note that the return currents are the same as the critical currents at $T = 250$ mK, indicating there is no Joule heating at this temperature. 

\begin{figure}
\centering
\includegraphics[width=8cm]{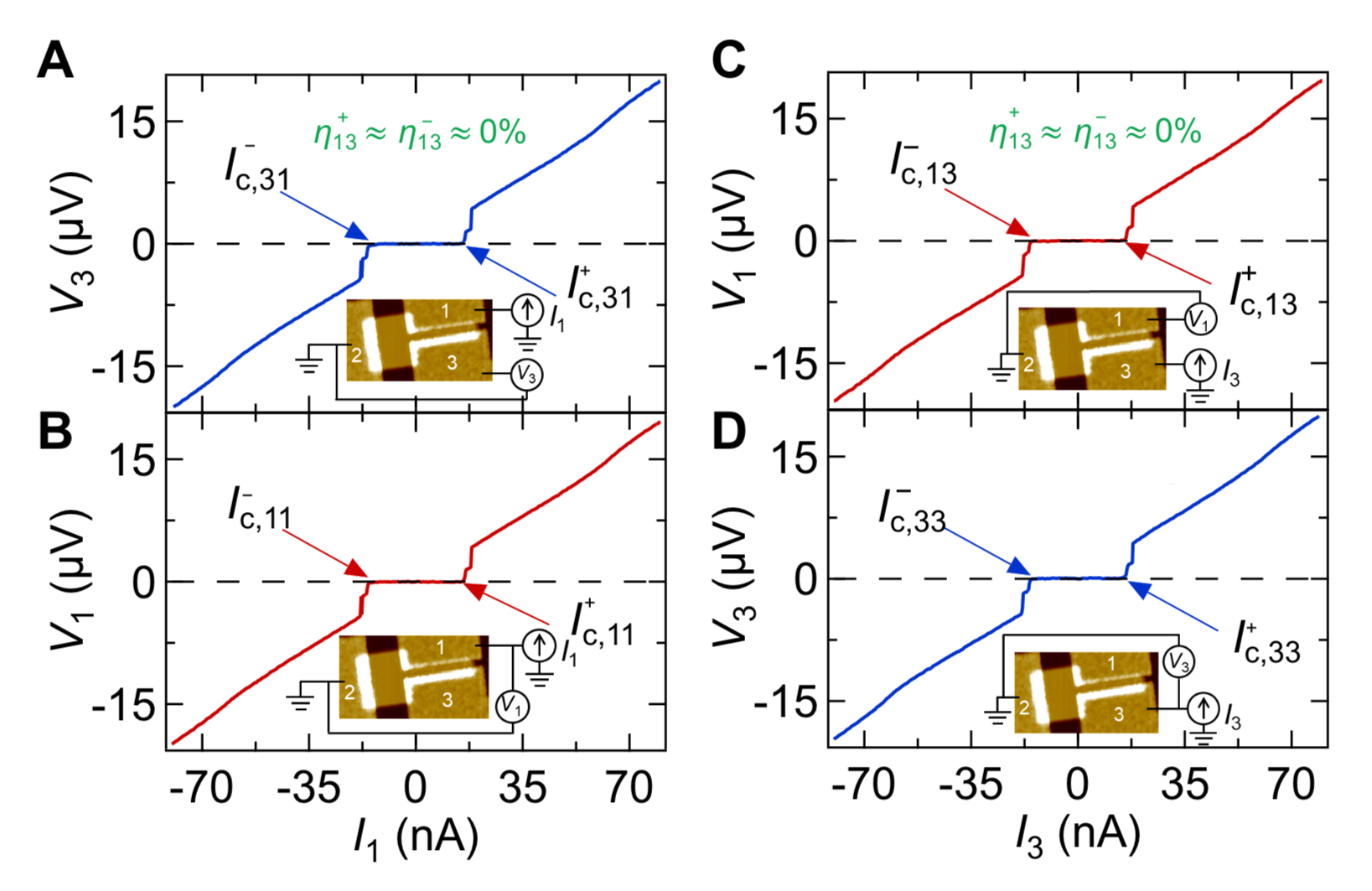}
\caption{\label{Fig_IV_symmetric.jpg} 
Voltage-current characteristics of the three-terminal JJ in Config.~1. 
(\textbf{A} and \textbf{B}) Voltages $V_{3}$ (\textbf{A}) and $V_1$ (\textbf{B}) as a function of the bias current $I_{1}$ at $I_{3} =$ 0 nA. 
(\textbf{C} and \textbf{D}) Voltages $V_{1}$ (\textbf{C}) and $V_3$ (\textbf{D}) as a function of the bias current $I_{3}$ at $I_{1}$  = 0 nA.
The inset in each panel shows the measurement configuration. In each measurement, $I_{1}$ (or $I_{3}$) is swept from $0$ nA to $100$ nA in the positive direction and from $0$ nA to $-100$ nA in the negative direction. 
The critical currents are labeled as $I_{c,13}^{\pm}$, $I_{c,31}^{\pm}$, $I_{c,11}^{\pm}$, and $I_{c,33}^{\pm}$.
}
\end{figure}

The asymmetry of the critical current for positive and negative polarities ($I_c^+ \neq I_c^-$) has been used to define the SDE in single-port networks (two-terminal devices) \cite{ando2020observation,diez2021magnetic,shin2021magnetic,baumgartner2022supercurrent,narita2022field,jeon2022zero,bauriedl2022supercurrent,pal2022josephson,wu2022field,bauriedl2022supercurrent,lin2022zero,matsuo2023josephson}. In MTJJs, this asymmetry depends on the bias current $I_3$ and has been considered as a signature of the SDE with a diode efficiency of  $\sim 100\%$ \cite{chiles2022non}. However, this single-port definition of nonreciprocity is not applicable in multi-port networks \cite{matsuo2023josephson}.
This is because the additional bias current $I_3$ may partially cancel out the current in terminal 1 ($I_1$), thereby shifting the critical current of this terminal toward either a positive or negative polarity (i.e., $I_{c1}^+ \neq I_{c1}^-$). Therefore, even for a reciprocal junction, this single-port definition of nonreciprocity may give an erroneous impression that transport is nonreciprocal. The limitations of this single-port definition of nonreciprocity are better showcased in the thought experiment discussed in Appendix ~\ref{thought experiment}. In the following, we will evaluate our MTJJs using the multiport definition of nonreciprocity.

We consider MTJJs as multiport networks and accordingly define the efficiency tensor for nonreciprocal transport as follow:
\begin{equation}\label{nr}
\eta_{jk}^\vartheta=\frac{|I_{c,jk}^\vartheta|-|I_{c,kj}^\vartheta|}{|I_{c,jk}^\vartheta|+|I_{c,kj}^\vartheta|},
\end{equation}
where $\vartheta=\pm $ specifies the input current polarity and $I^\vartheta_{c,jk}$ denotes the critical current tensor which is defined as the maximum/minimum current $I_k$ for which the voltage at terminal $j$ remains zero: 
\begin{equation}\label{nrc}
I_{c,jk}^{+}=\text{max}\big[I_k\big]_{V_j=0}, I_{c,jk}^{-}=\text{min}\big[I_k\big]_{V_j=0} ,  I_\ell=\text{0},  \forall \ell \neq j ,
\end{equation}

We first focus on the three-terminal JJ in Confing. 1, where terminal 2 is grounded and a two-port network is formed between terminals 1 and 3.
Figure~\ref{Fig_IV_symmetric.jpg} plots the voltage-current characteristics of the three-terminal JJ. Following Eq.~\ref{nrc}, we extract the critical currents $I_{c,11}^{\pm}$, $I_{c,31}^{\pm}$, $I_{c,13}^{\pm}$, and $I_{c,33}^{\pm}$ which are marked by arrows. We note that these critical currents are  independent of the polarity of the applied current, i.e.,  $\vert I_{c,jk}^+ \vert = \vert I_{c,jk}^- \vert$  where $j, k = 1, 3$. We find that $|I_{c,11}^{\pm}| = |I_{c,13}^{\pm}| = |I_{c,31}^{\pm}| = |I_{c,33}^{\pm}| = 15.75$ nA, indicating that transport is reciprocal in this configuration. 

We explain this observation by considering the circuit-network of coupled JJs shown in the inset of Fig.~\ref{P1.png}B. In Figs.~\ref{Fig_IV_symmetric.jpg}A and B, $I_1$ splits between JJ1, and JJ2 in series with JJ3. Since terminals 1 and 3 are mirror symmetric ($I_{c1} \approx I_{c2}$), JJ1 and JJ2 transition to the normal state simultaneously once $I_1 > I_{c1}+I_{c2} \approx 2I_{c1}$. As a result, a non-zero voltage develops across terminals 1 ($V_1 \neq 0$) and 3 ($V_3 \neq 0$). Following Eq.~\ref{nrc}, we find $I_{c,11}^+ \approx I_{c,31}^+ \approx 2I_{c1}$. Likewise, in Figs.~\ref{Fig_IV_symmetric.jpg}C and D, $I_3$ splits between JJ2, and JJ1 in series with JJ3. Similarly in this case, JJ1 and JJ2 transition to the normal state once $I_3 > I_{c1}+I_{c2} \approx 2I_{c1}$. Following Eq.~\ref{nrc}, we find that $I_{c,13}^+ \approx I_{c,33}^+ \approx 2I_{c1}$. Therefore, all the critical currents  are the same in this configuration. Considering $I_{c,11}^{+} \approx 2I_{c1} \approx 15.75$ nA, we estimate $I_{c1} \approx I_{c2} \approx 7.88$ nA.
\begin{figure}
\centering
\includegraphics[width=8cm]{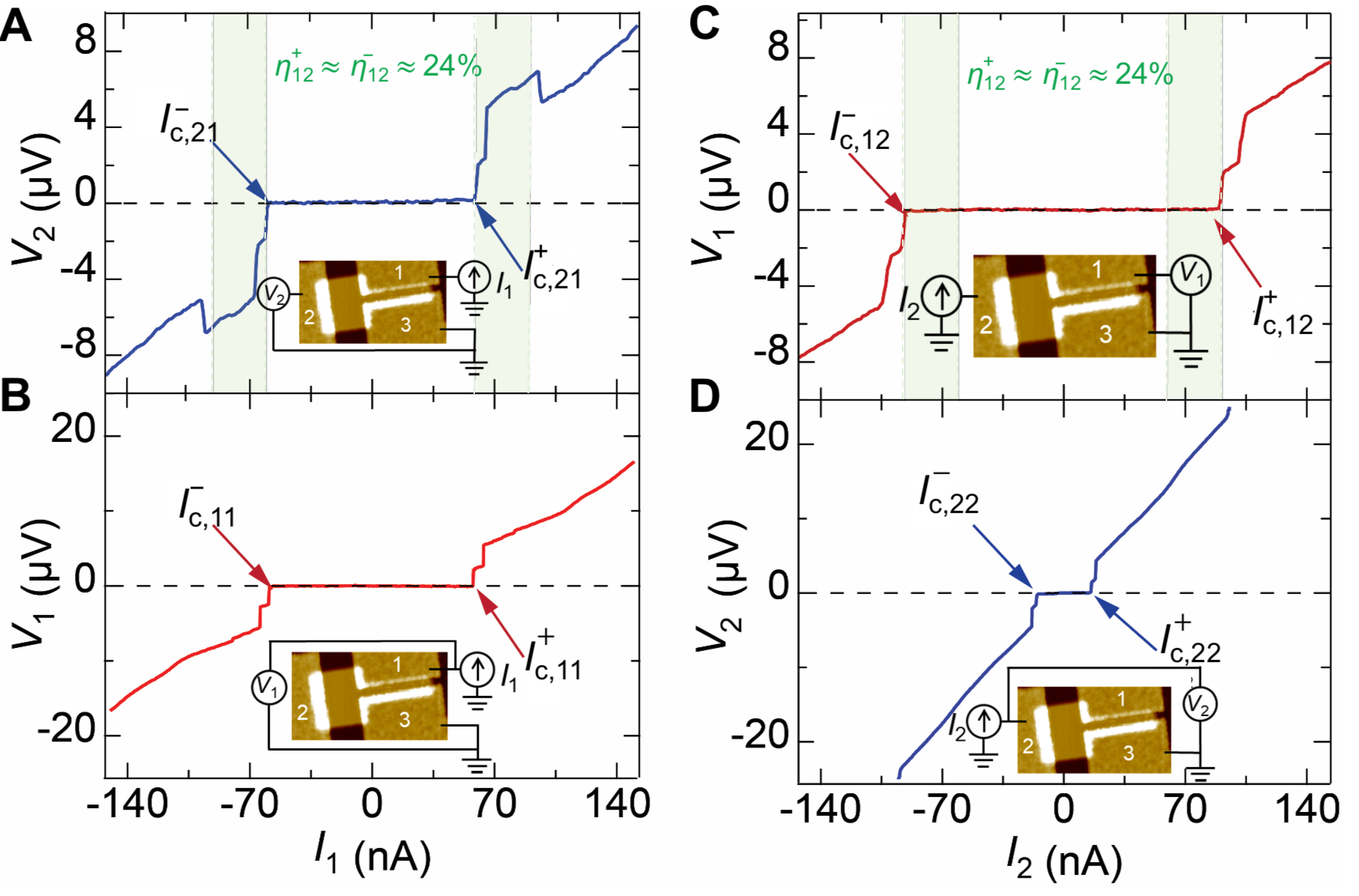}
\caption{\label{Fig_IV_asymmetric.jpg} 
Voltage-current characteristics of the three-terminal JJ in Config.~2. 
(\textbf{A} and \textbf{B}) Voltages $V_{2}$ (\textbf{A}) and $V_1$ (\textbf{B}) as a function of the bias current $I_{1}$ at $I_{2} =$ 0 nA. 
(\textbf{C} and \textbf{D}) Voltages $V_{1}$ (\textbf{C}) and $V_2$ (\textbf{D}) as a function of the bias current $I_{2}$ at $I_{1}$  = 0 nA.
The inset in each panel shows the measurement configuration. In each measurement, $I_{1}$ (or $I_{2}$) is swept from $0$ nA to $150$ nA in the positive direction and from $0$ nA to $-150$ nA in the negative direction. 
The critical currents are labeled as $I_{c,12}^{\pm}$, $I_{c,21}^{\pm}$, $I_{c,11}^{\pm}$, and $I_{c,22}^{\pm}$.
Shaded areas in panels (\textbf{A}) and (\textbf{C})  mark the range of the current in which transport is nonreciprocal. The efficiency of  two-port nonreciprocity $\eta^{\pm}_{12}=(|I^{\pm}_{c,12}|-|I^{\pm}_{c,21}|)/(|I^{\pm}_{c,12}|+|I^{\pm}_{c,21}|)$ are calculated from the critical currents as $\eta^+_{12} \approx \eta^-_{12}  \approx 24\%$.
}
\end{figure}
We now investigate the transport properties of the three-terminal JJ in Confing. 2, where terminal 3 is grounded and a two-port network is formed between terminals 1 and 2 (See Appendix ~\ref{dIdVmap} for differential resistance maps in this configuration). Figure~\ref{Fig_IV_asymmetric.jpg} plots the voltage-current characteristics of the three-terminal JJ in Config. 2, where the critical currents $I_{c,11}^{\pm}$, $I_{c,21}^{\pm}$, $I_{c,12}^{\pm}$, and $I_{c,22}^{\pm}$ are marked by arrows. Similar to Config. 1, the critical currents are  identical for both polarities of the applied current, i.e.,  $\vert I_{c,jk}^+ \vert = \vert I_{c,jk}^- \vert$ where $j, k = 1, 2$. Notably, we find that $I_{c,12}^{\pm} \neq I_{c,21}^{\pm}$, indicating that transport is nonreciprocal within the shaded regions in Figs.~\ref{Fig_IV_asymmetric.jpg}A and C. Table~\ref{Table} summarizes the extracted values of the critical current for both configurations.      
 \begin{table}
\centering
\begin{tabular}{|c|c|c|}
\hline
\parbox[t]{15mm}{\multirow{4}{*}{\rotatebox[origin=c]{0}{Config.~1}}}   & \multicolumn{1}{c|}{$|I_{c,11}^{\pm}|$ }   & 15.75 nA \\ \cline{2-3}
                                                                       &                     $|I_{c,13}^{\pm}|$    & 15.75 nA \\ \cline{2-3}
                                                                       &                     $|I_{c,31}^{\pm}|$    & 15.75  nA\\ \cline{2-3}
                                                                       &                     $|I_{c,33}^{\pm}|$    & 15.75  nA\\ \hline
\parbox[t]{15mm}{\multirow{4}{*}{\rotatebox[origin=c]{0}{Config.~2}}}   & \multicolumn{1}{c|}{$|I_{c,11}^{\pm}|$}   & 57.5 nA \\ \cline{2-3}
                                                                       &                     $|I_{c,12}^{\pm}|$    & 94.25  nA\\  \cline{2-3}
                                                                       &                     $|I_{c,21}^{\pm}|$    & 57.25  nA\\ \cline{2-3}
                                                                       &                     $|I_{c,22}^{\pm}|$    & 15.75 nA \\ \hline
\end{tabular}

\caption{The critical current tensor measured in both configurations.}
\label{Table}%
\end{table}

The observed nonreciprocity in Config. 2 arises from the broken spatial mirror symmetry  between terminals 1 and 2. We explain this nonreciprocity by considering  the circuit-network of the coupled JJs as shown in the inset of Fig.~\ref{P1.png}B. When the current $I_1$ is applied to terminal 1 but terminal 2 is floating (Figs.~\ref{Fig_IV_asymmetric.jpg}A and B), $I_1$ splits between JJ3, and JJ1 in series with JJ2. Once $I_1 > I_{c3} + I_{c1}$, all three JJs simultaneously transition to the normal state, resulting in nonzero $V_1$ and $V_2$. From Eq.~\ref{nrc}, we estimate $I_{c,11}^{+} \approx I_{c,21}^{+} \approx I_{c3} + I_{c1}$. We also experimentally extract $I_{c,11}^{+} \approx I_{c,21}^{+} \approx  57.25$ nA (Table~\ref{Table}). Therefore, $I_{c3} + I_{c1} \approx 57.25$ nA. Considering the value of $I_{c1} \approx I_{c2} = 7.88$ nA obtained in Config. 1, we estimate $I_{c3} \approx 49.4$ nA.

On the other hand, when the current $I_2$ is applied to terminal 2 but terminal 1 is floating (Figs.~\ref{Fig_IV_asymmetric.jpg}C and D), $I_2$ splits between JJ2, and JJ1 in series with JJ3. In this case, when $I_2 > I_{c1}+I_{c2} \approx 2I_{c1}$, JJ1 and JJ2 simultaneously transition to the normal state, resulting in nonzero $V_2$. Following Eq.~\ref{nrc}, we estimate $I_{c,22}^+ \approx 2I_{c1} \approx 15.75$ nA, which is consistent with $I_{c,22}^+ \approx 15.75$ nA that we obtain experimentally (Table~\ref{Table}). However, because $I_{c3} \approx 49.4$ nA $>I_{c1}+I_{c2} \approx 15.75$ nA, JJ3  remains in the superconducting state, resulting in $V_1 = 0$.
The current flow in JJ3 depends on the ratio of the normal state resistances between JJ1 ($R_{1}$) and JJ2 ($R_{2}$), i.e., $I_{JJ3} \approx \frac{R_{2}}{R_{2}+R_{1}}I_2$. Assuming $R_1 \approx R_2$, $I_2 \approx 2I_{JJ3}$. Therefore, when $I_{JJ3} > I_{c3}$ or $I_2 > 2I_{c3}$, JJ3 transitions to the normal state, resulting in a nonzero $V_1$. From this analysis, we estimate $I_{c,12}^+ = 2I_{c3} \approx 98.8$ nA, which is consistent with $I_{c,12}^+ = 94.25$ nA that we obtain experimentally (Table~\ref{Table}). We further calculate the nonreciprocal efficiency of this two-port network (Eq.~\ref{nr}) as $\eta_{12}^{+} \approx \eta_{12}^- \approx  24\%$ (see Appendix ~\ref{efficiency limit} for results from a more symmetric device). We finally note that, based on our above analysis, this efficiency is maximized when $I_{c3} \gg I_{c1}\approx I_{c2}$. In this case, $I_{c,12} = 2I_{c3}$ and $I_{c,21} = I_{c3} + I_{c1} \approx I_{c3}$, resulting in $\eta^\pm_{12} \approx 33 \%$.

To better understand the limits of the efficiency, we have performed numerical simulations of our circuit-network model using resistively shunted junctions (RSJs). In our simulations, we consider the ratio between the critical currents of JJ3 and JJ1/JJ2 as a representation of the spatial mirror symmetry breaking (i.e., $x = I_{c3}/I_{c1}=I_{c3}/I_{c2}$). In this case, $x=1$ represents the perfect mirror symmetry and results in zero efficiency. Our modeling shows that the efficiency increases monotonically with increasing $x$ and eventually reaches a theoretical limit of 30\% (see Appendix ~\ref{efficiency limit} for more details).
Finally, we note that because there are both dissipative (JJ1 and JJ2) and superconducting (JJ3) channels during the measurement of $I_{c,12}^\pm$, the observed nonreciprocity in our three-terminal JJ is different from the purely superconducting SDE observed in two-terminal JJs. 
Our combined results from Configs. 1 and 2 indicate that two-port nonreciprocity can be engineered in MTJJs by changing the device geometry and/or measurement configurations. 
\begin{figure}
 \centering
 \includegraphics[width=8cm]{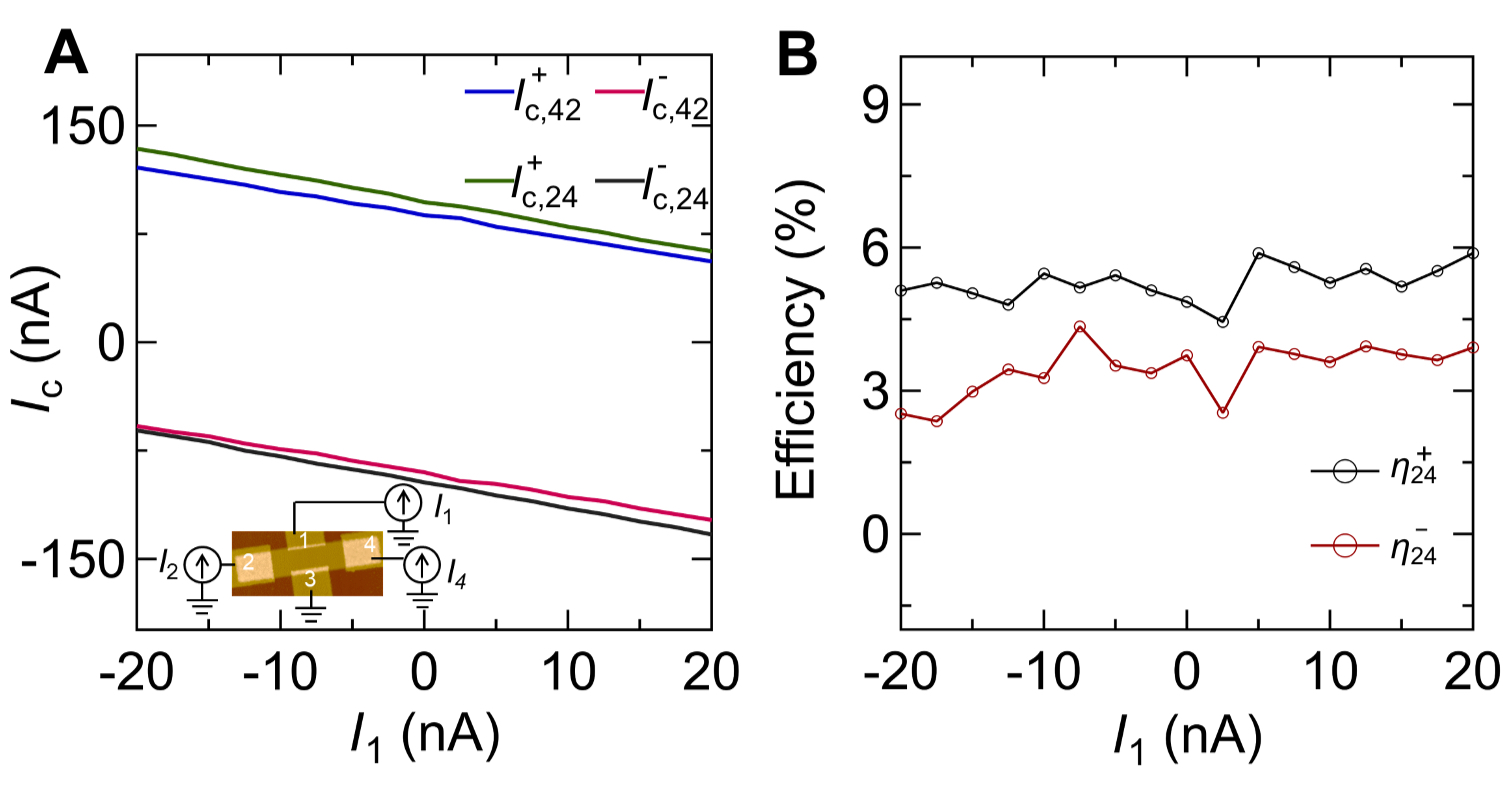}
\caption{\label{I1_I4.png}(\textbf{A}) Critical currents versus the bias current $I_{1}$ extracted from V-I characteristics of a four-terminal JJ. Inset shows the measurement configuration. Here, terminal 3 is grounded and $I_{1}$ is kept at a constant value during each measurement.  
(\textbf{B}) Efficiency of nonreciprocal transport $\eta^{\pm}_{24}$ as a function of $I_{1}$.
}
\end{figure}

To investigate the role of a control bias current on the two-port non-reciprocity, we study an asymmetric four-terminal JJ. Figure~\ref{I1_I4.png}A plots $I_{c,24}^{\pm}$ and $I_{c,42}^{\pm}$ as a function of $I_{1}$. Inset of Fig.~\ref{I1_I4.png}A shows the AFM image of the device as well as the measurement configuration. We ground terminal 3 and apply currents $I_2$ and $I_4$ while keeping $I_1$ constant. In this case, we can treat the four-terminal JJ as a three-port network (between terminals 1, 2, and 4). Following Eq.~\ref{nrc}, we extract the critical current tensor at zero and nonzero values of $I_1$. 
 Figure~\ref{I1_I4.png}B plots the efficiency of nonreciprocal transport as a function of $I_1$. 
 Because terminals 2 and 4 are mirror symmetric, we would expect no nonreciprocal transport. However, $\eta_{24}^{+} \approx \eta_{24}^- \approx 4 \%$. We speculate that because of the imperfect fabrication process, the contact transparency is slightly different between terminals 2 and 4, resulting in nonzero but negligible $\eta_{24}^{\pm}$. 

 
 From Fig.~\ref{I1_I4.png}A, we observe that $I_{c,24}^\pm$ and $I_{c,42}^\pm$ change almost linearly with $I_1$. However, the difference between $I_{c,24}^{\pm}$ and $I_{c,42}^{\pm}$ does not change significantly with $I_1$. Therefore, $\eta_{24}^+$ and $\eta_{24}^-$ remain independent of $I_1$. 
 We note that the geometry of the four-terminal JJ (inset of Fig.~\ref{I1_I4.png}A) is such that terminals 2 and 4 are mirror symmetric even for $I_1 \neq 0$. 
 Therefore, our observations indicate that the effect of the control current ($I_1$) on two-port nonreciprocity is negligible compared to the spatial mirror symmetry of the terminals (please see Appendix ~\ref{4TJJ} for additional measurements between terminals 1 and 2).

\section{Conclusion}

In conclusion, we have studied multi-port nonreciprocal transport in MTJJs. Following the reciprocal relation in two-port electrical networks, we have developed a definition of two-port nonreciprocity for MTJJs. We have also explored transport properties of three- and four-terminal JJs with asymmetric graphene channels. We have shown that the MTJJ exhibits magnetic-field-free nonreciprocal (reciprocal) transport if spatial mirror symmetry is broken (preserved) between the terminals. We have achieved these symmetry conditions by designing our MTJJs such that the graphene channel is symmetric between some but not all of the terminals. This design has allowed us to engineer reconfigurable nonreciprocal properties in our MTJJs. We have further studied the impact of a control bias current on the efficiency of nonreciprocal transport. We have also demonstrated nonreciprocity that is independent of the bias current polarity. This polarity-independent two-port nonreciprocity might potentially be useful for quantum applications where directional transport for drive and read-out microwave signals are required. Overall, our findings demonstrate the potential for exploiting MTJJs as a reconfigurable platform to engineer magnetic-field-free multi-port devices for applications in low-temperature superconducting logics \cite{golod2022demonstration}  and directional cryogenic devices \cite{leroux2022nonreciprocal}. 

\begin{acknowledgements}
We acknowledge funding from the National Science Foundation (NSF) Innovation and Technology Ecosystems (No. 2040667). F.Z. and N.S. acknowledge support from the University of Chicago. G.J.C. acknowledges support from the ARAP program of the Office of the Secretary of Defense. M.J.G. and M.T.A. acknowledge funding from US ARO Grant W911NF-20-2-0151 and the NSF through the University of Illinois at Urbana-Champaign Materials Research Science and Engineering Center DMR-1720633. K.W. and T.T. acknowledge support from the JSPS KAKENHI (Grant Numbers 19H05790, 20H00354 and 21H05233). 
\end{acknowledgements}

\noindent\textbf{Data Availability:}
The data supporting the conclusions of this letter is available on Zenodo (https://doi.org/10.5281/zenodo.8066318).

\appendix
\section{Stochastic switching noise at T = 12 mK} \label{Stochastic switching}

\begin{figure}[h]
 \centering
 \includegraphics[width=8cm]{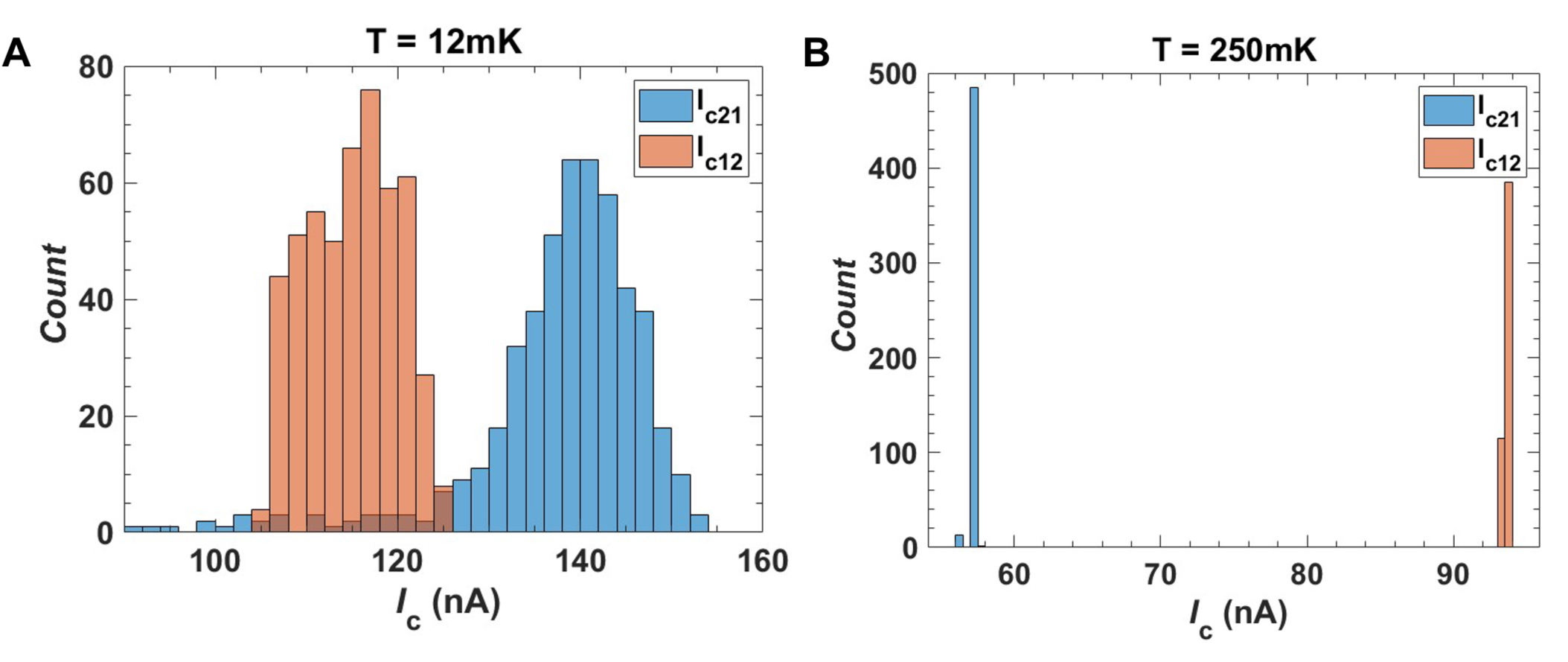}
\caption{\label{S1.jpg} (\textbf{A, B}) Histogram of critical currents $I_{c,12}^+$ and $I_{c,21}^+$ in the three-terminal JJ of the main text, measured at T = 12 mK (\textbf{A}) and T = 250 mK (\textbf{B}).}
\end{figure}

Figure~\ref{S1.jpg} plots the histogram of the critical currents $I_{c,12}^{+}$  and $I_{c,21}^{+}$  at two different temperatures: T = 12 mK (A) and T = 250 mK (B) for the three-terminal JJ of the main text. We observe that at T = 12 mK, the device exhibits significant stochastic switching noise. However, the switching noise almost vanishes at T = 250 mK.

\section{Differential resistance maps from the three-terminal JJ in Config. 2.} \label{dIdVmap}

\begin{figure}[h]
 \centering
 \includegraphics[width=8cm]{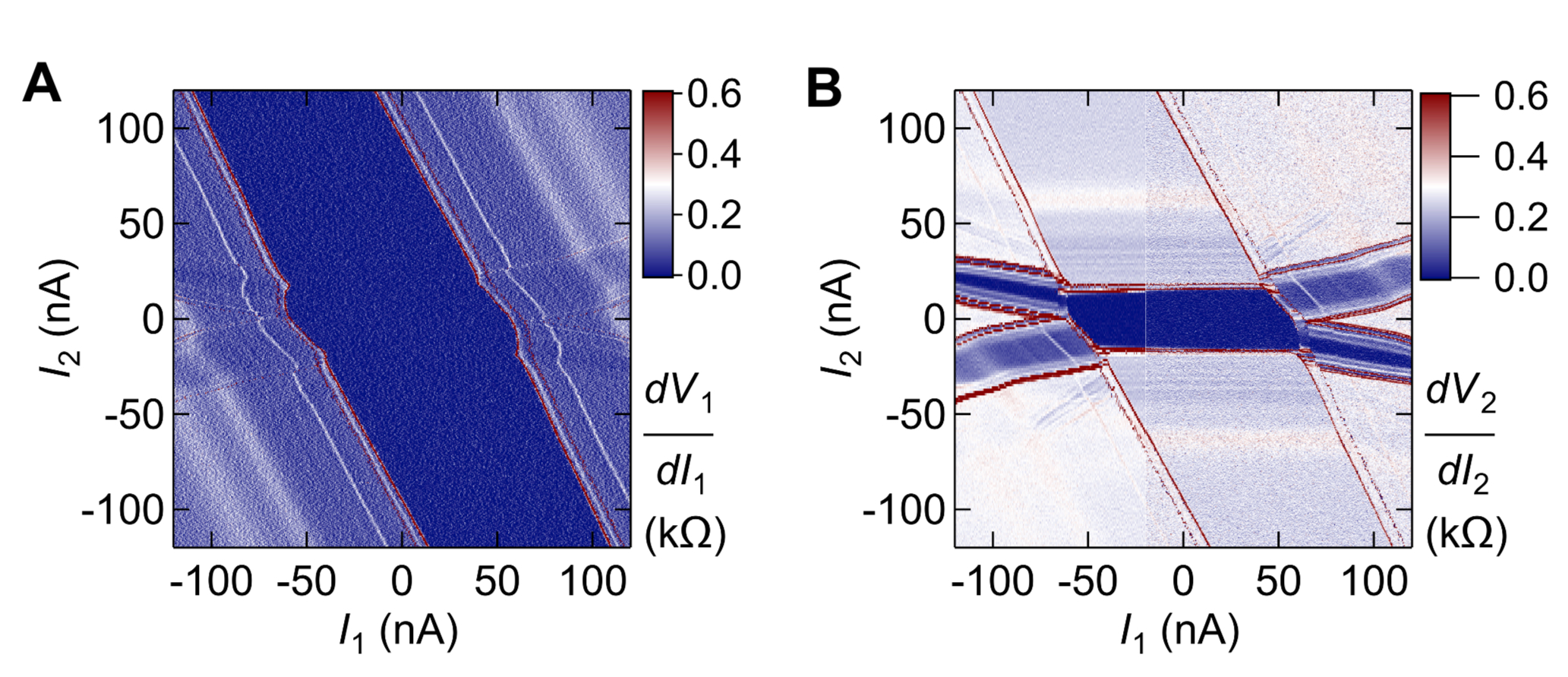}
\caption{\label{S2.jpg} 
Differential resistance maps measured in the three-terminal JJ of the main text in Config. 2. (\textbf{A,B}) Color map of the differential resistance $dV_1/dI_1$(\textbf{A}) and  $dV_2dI_2$ (\textbf{B}) versus $I_1$ and $I_2$ in Config 2.}
\end{figure}

Figure~\ref{S2.jpg} 
plots the differential resistance $dV_{1}/dI_{1}$ and $dV_{2}/dI_{2}$ maps versus $I_{1}$ and $I_{2}$ for the three-terminal JJ of the main text in Config. 2. In panel A (B), we sweep $I_{1}$ ($I_{2}$) for each fixed $I_{2}$ ($I_{1}$) value. In this configuration, terminal 3 is grounded. 

\section{Nonreciprocal transport in a second asymmetric three-terminal JJ} \label{Nonreciprocal 3TJJ}
\begin{figure}[h]
 \centering
 \includegraphics[width=8cm]{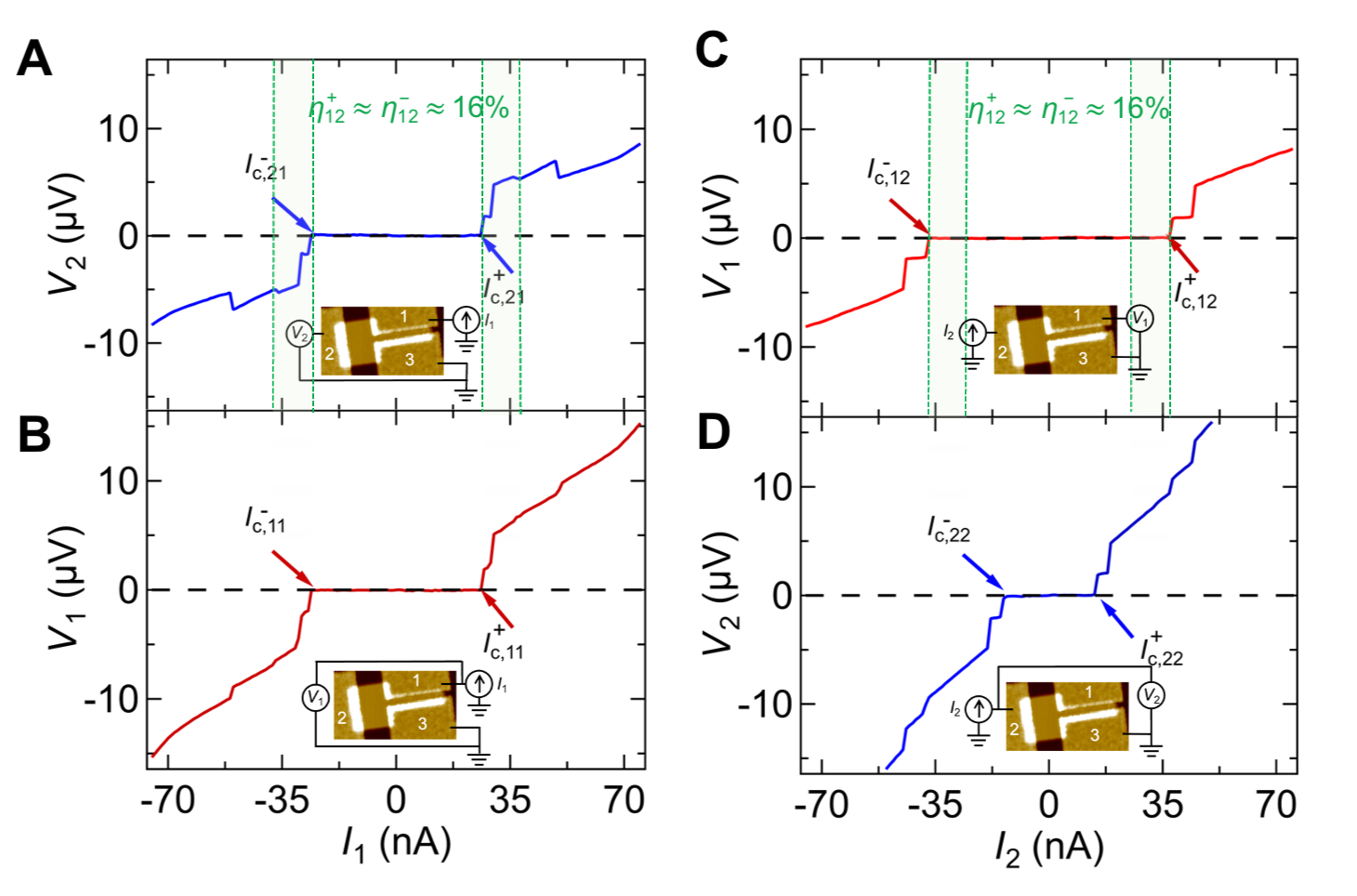}
\caption{\label{S3.jpg} 
(\textbf{A,B}) Voltages $V_{2}$ (\textbf{A}) and $V_1$ (\textbf{B}) as a function of the bias current $I_{1}$ at $I_{2} =$ 0 nA. 
(\textbf{C,D}) Voltages $V_{1}$ (\textbf{C}) and $V_2$ (\textbf{D}) as a function of the bias current $I_{2}$ at $I_{1}$  = 0 nA.
The inset in each panel shows the measurement configuration. In each measurement, $I_{1}$ (or $I_{2}$) is swept from $0$ nA to $100$ nA in the positive direction and from $0$ nA to $-100$ nA in the negative direction. 
The critical currents are labeled as $I_{c,11}^{\pm}$, $I_{c,21}^{\pm}$, $I_{c,12}^{\pm}$, and $I_{c,22}^{\pm}$. Shaded areas in panels (\textbf{A}) and (\textbf{C})  mark the range of the current in which transport is nonreciprocal. The efficiency of  two-port nonreciprocity are calculated as $\eta^+_{12} \approx \eta^-_{12}  \approx 16\%$.
}
\end{figure}

Figure ~\ref{S3.jpg} plots the voltage-current characteristics of another asymmetric three-terminal JJ in Config. 2. Here, we also observe nonreciprocal transport between terminals 1 and 2 which have no mirror symmetry. The efficiency is $\eta^+_{12} \approx 16\%$. The inset in each panel shows the bias configuration. 

\section{Nonreciprocal transport in the four-terminal JJ }\label{4TJJ}
\begin{figure} [h]
 \centering
 \includegraphics[width=8cm]{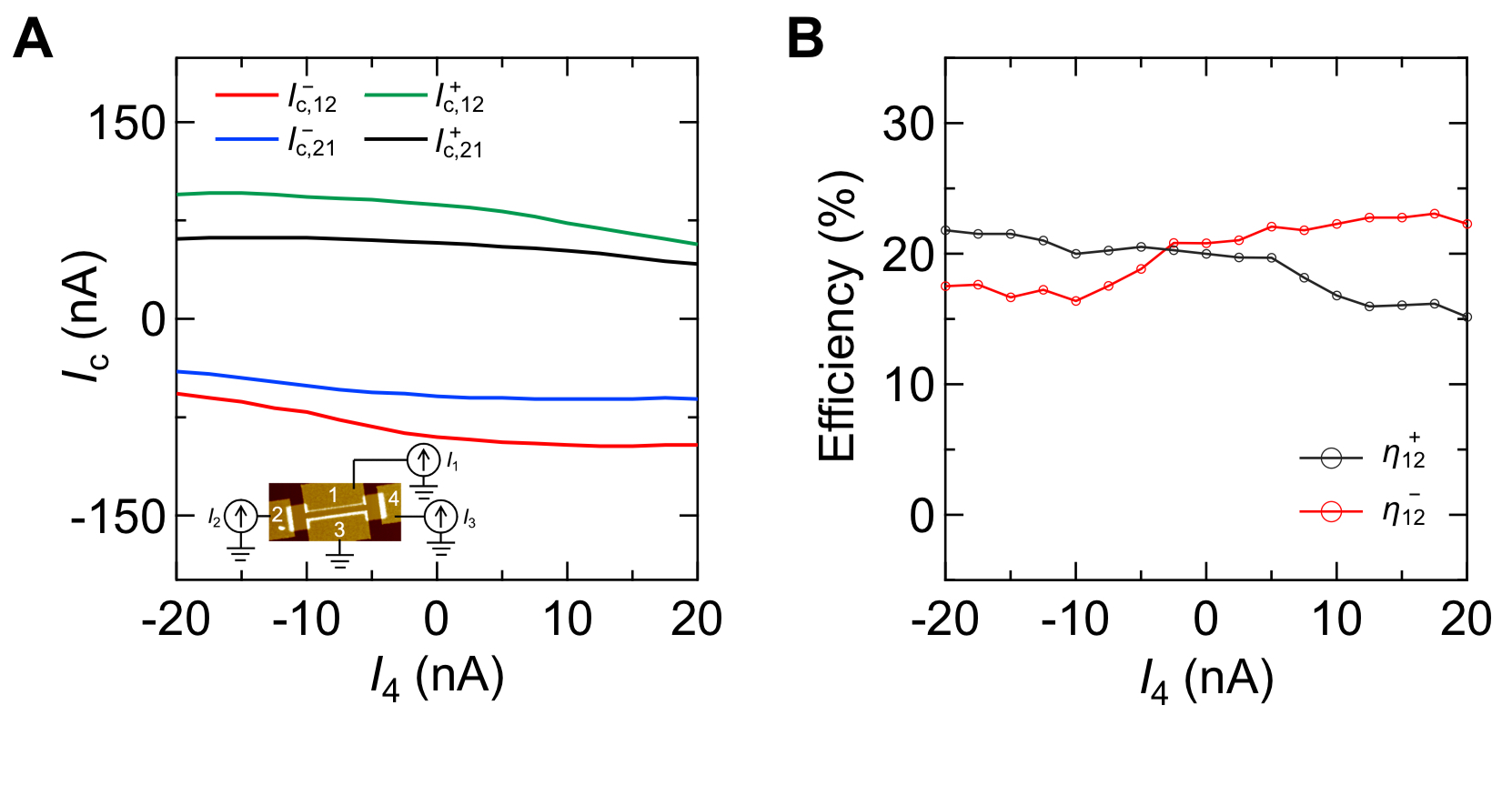}
\caption{\label{S4.jpg} 
Nonreciprocal transport in the four-terminal JJ of the main text, measured between terminals 1 and 2. (\textbf{A}) Critical currents versus the bias current $I_4$ extracted from V-I characteristics of a four-terminal JJ. Inset shows the measurement configuration. Here, terminal 3 is grounded and $I_4$ is kept at a constant value during each measurement. (\textbf{B}) Efficiency of nonreciprocal transport $\eta^{\pm}_{12}$ as a function of $I_4$.
}
\end{figure}

Figures~\ref{S4.jpg}A and B plot the critical currents $I_{c,12}^{\pm}$ and $I_{c,21}^{\pm}$ (A) and efficiency $\eta^+_{12}$  as a function of the control current $I_{4}$ in the four-terminal JJ of the main text. The inset of Fig.~\ref{S4.jpg}A shows the bias configuration. At $I_{4} = 0$, we observe that the critical current $I_{c,12}^{\pm} \neq I_{c,21}^{\pm}$, indicating that transport between these two terminals is nonreciprocal. Similar to the three-terminal case, this nonreciprocity is due to the asymmetric geometry of the graphene channel. We further observe that a nonzero $I_{4}$ creates an imbalance between positive and negative critical current polarities. We observe that while nonreciprocal transport is polarity independent for $I_{4} = 0$, it becomes polarity dependent for $I_{4}\neq 0$, i.e., $\eta^+_{12} \neq \eta^+_{21}$. However, we note that the average value of $\eta^{avg}_{12} = \frac{(\eta^+_{12}+\eta^-_{12} )}{2}\approx 20\%$ remains approximately constant. Therefore, we conclude that mirror symmetry breaking is responsible for nonreciprocal transport, whereas a control bias current simply creates a polarity-dependent imbalance between the critical currents.

\section{Single-port treatment of MTJJs and asymmetric critical currents}\label{thought experiment}

\begin{figure} [h]
 \centering
 \includegraphics[width=4cm]{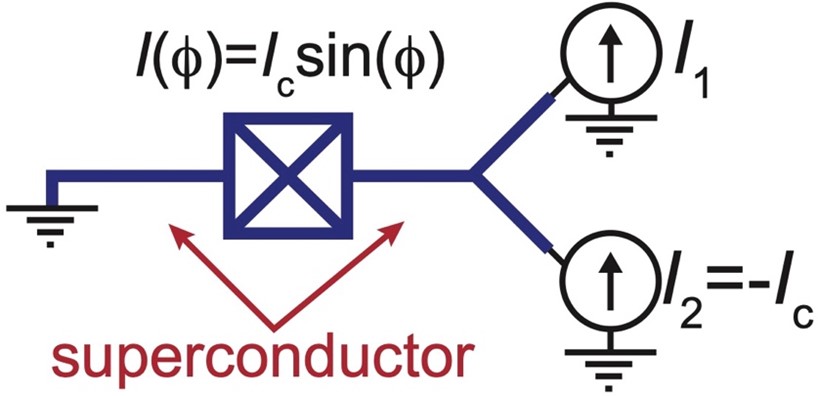}
\caption{\label{S5.jpg} 
Schematic of a two-terminal JJ connected to two current sources. The current in the JJ ($I_{JJ}$) is the sum of $I_1$ and $I_2$. If $I_2=-I_c$, where $I_c$ is the critical current of the JJ, $I_{JJ}=I_1-I_c$. In this setup, the JJ is in the superconducting state for $0\le I_1\le 2I_c$, resulting in $I^+_{c,11}=2I_c$ and $I_{c,11}^-$ = 0. Therefore, if $I_{c,11}$ is solely used to characterize the JJ, one might assume the device is nonreciprocal, even though the JJ itself is reciprocal. This toy example highlights the limitations of using the single-port definition of nonreciprocity in multi-port networks.
}
\end{figure}

In the three-terminal JJ of the main text, we observe that for nonzero $I_{2}$, the critical current at terminal 1 ($I_{c,11}$) is asymmetric for positive and negative current polarities (Fig.~\ref{P1.png}D of the main text). This asymmetry may be interpreted as a signature of nonreciprocal transport in three-terminal JJs. However, this interpretation is based on a single-port definition of the critical current, and as we argue in the following is not applicable in multi-port devices. In the schematic of Figure ~\ref{S5.jpg}, we consider a reciprocal two-terminal JJ with a critical current $I_{c}$. We further assume that the left terminal is grounded but the right terminal is connected through superconducting leads to two current sources ($I_{1}$ and $I_{2}$). If $I_{2}=-I_{c}$, $I_{JJ}=I_{1}+I_{2}=I_{1}-I_{c}$. In this case, when $I_{1}=2I_{c}$, $I_{JJ}=I_{1}-I_{c}=I_{c}$ and the JJ will transition to normal state. Following the single-port definition of nonreciprocity, this value of $I_{1}$ is considered $I^+_{c1}=2I_{c}$. On the other hand, when $I_1=0$, $I_{JJ}=I_1-I_c=-I_c$. This value is then considered $I^-_{c1}$. Consequently, in this three-terminal system, $I^+_{c1} \neq I^-_{c1} $. However, as we noted earlier the two-terminal JJ is reciprocal. We note that the use of superconducting leads ensures that the current injected into the JJ is dissipationless. Therefore, even though the underlying current-phase relation of the three-terminal JJ may be asymmetric with the sign reversal of the phase at $I_2 \neq 0$, [$I_1 (\Phi_1,I_2 ) \neq -I_1 (-\Phi_1,I_2)$], it is experimentally challenging to distinguish reciprocal and nonreciprocal devices using the single-port definition of nonreciprocity. 
\section{The resistively shunted junction (RSJ) model parameters}\label{RSJ}
We use the RSJ model similar to our previous work \cite{Zhang2023Andreev} to numerically simulate our three-terminal junctions. Here, we assume the capacitance of the junction is zero. Following Ref. \cite{Zhang2023Andreev}, for our three-terminal device, we obtain:

\begin{equation}
   \boldsymbol{I}=\boldsymbol{f}(\tau)+\mathcal{G}\cdot\frac{d\boldsymbol{\Phi}(\tau)}{d\tau}.
\end{equation}
where $ \boldsymbol{I} = (I_1/I_c,I_2/I_c)^T$, $\boldsymbol{\Phi} = (\phi_1(\tau),\phi_2(\tau))^T$, and 
\begin{equation}
     \boldsymbol{f}(\tau)=\begin{pmatrix}
  I_{c3}\sin\phi_{13}+I_{c1}\sin\phi_{1}\\
   I_{c2}\sin\phi_{23}+I_{c1}\sin\phi_{2}
    \end{pmatrix}.
\end{equation}
In the above equations, $\tau = t/\tau_J$, where $t/\tau_J=\hbar/2eRI_{c}$ with $R = 1 $ $\Omega$ and $I_c$ = 10 nA. The parameters of our RSJ model are: $G_{12}=G_{23}=1/115  ~ (1/\Omega)$, $G_{13}=xG_{12}$ and $I_{c1}=I_{c2}=1.3, I_{c3}=xI_{c1}$.

\section{Dependence of nonreciprocal efficiency  on channel asymmetry}\label{efficiency limit}

\begin{figure}[h]
 \centering
 \includegraphics[width=7cm]{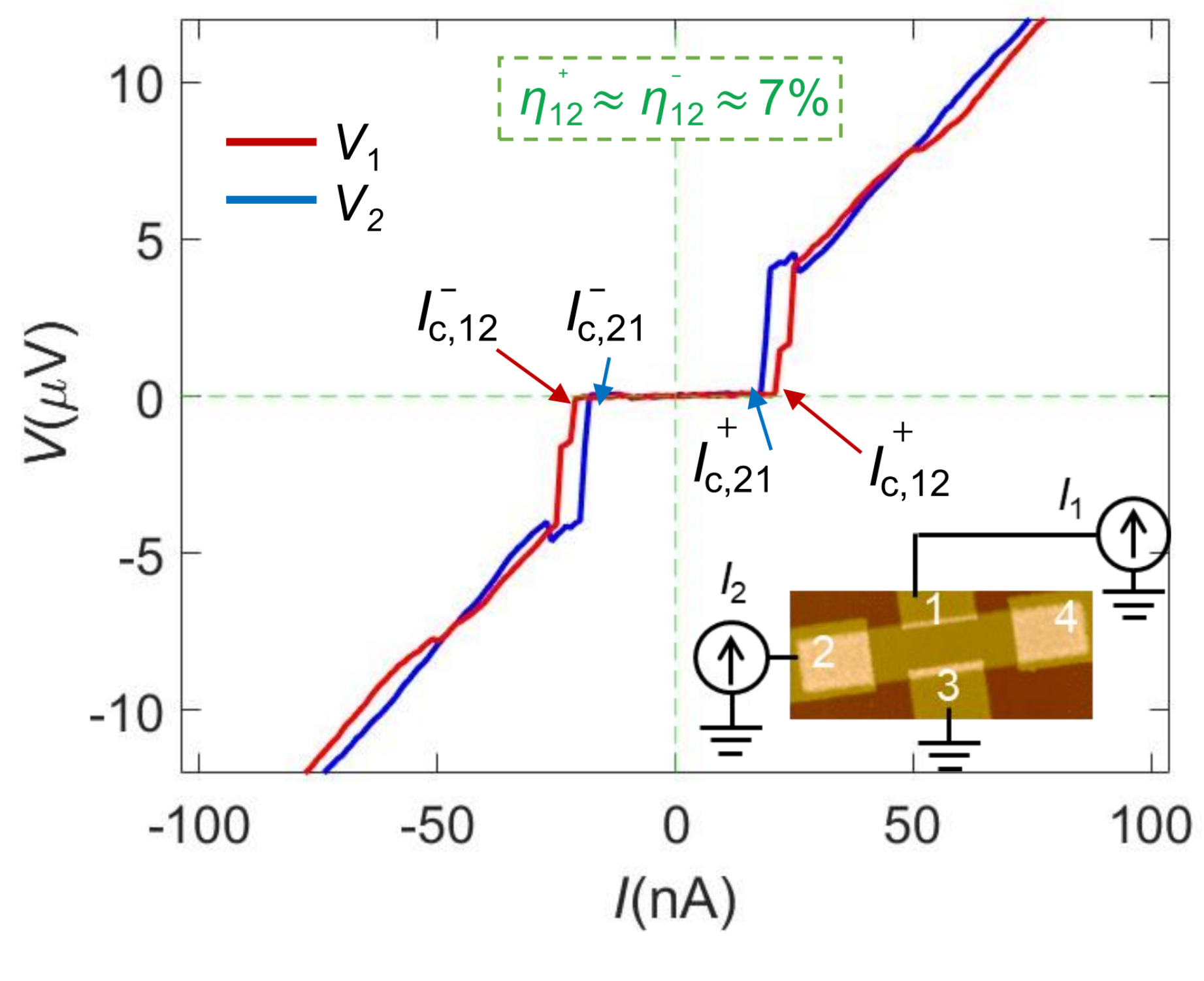}
\caption{\label{S6.jpg} 
Nonreciprocal transport in a second four-terminal JJ. Voltages $V_1$ and $V_2$ versus bias currents $I_2$ and $I_1$ at $I_1$ = 0 nA and $I_2$ = 0 nA, respectively. Arrows indicate the critical currents as $I_{c,12}^+, I_{c,21}^+, I_{c,12}^-,$ and  $I_{c,21}^-$. The efficiency of two-port nonreciprocity is $\eta^+_{12} \approx \eta^-_{12}  \approx 7\%$.
}
\end{figure}

\begin{figure}[h]
 \centering
 \includegraphics[width=7cm]{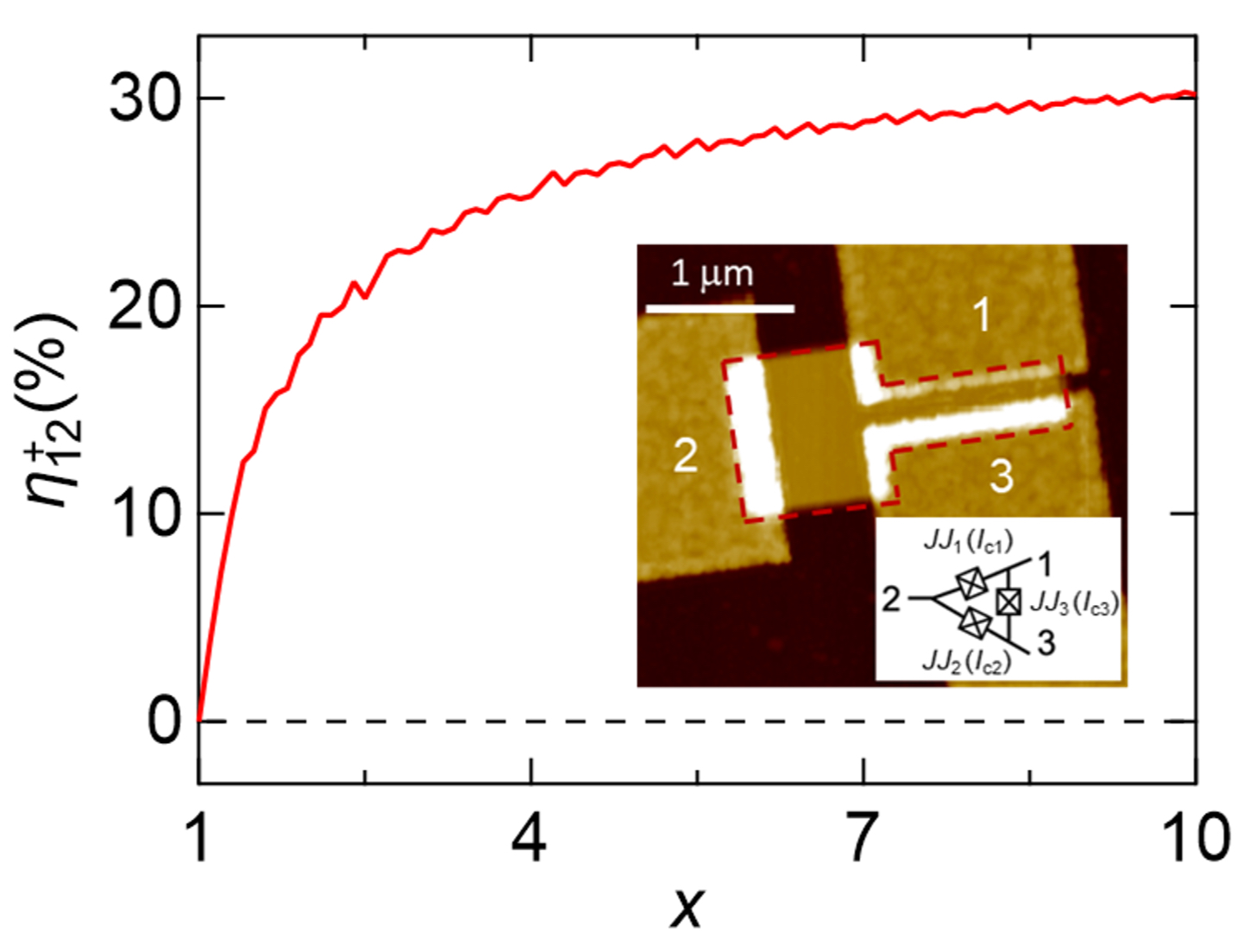}
\caption{\label{S7.jpg} 
Theoretical simulation of the efficiency $\eta^+_{12}$ using RSJ model. The value of $x$ represents the degree of asymmetry in the three-terminal JJ as $x=I_{c3}/I_{c1} = I_{c3}/I_{c2}$.
}
\end{figure}

Figure ~\ref{S6.jpg} plots the voltage-current characteristics of another four-terminal JJ, where the graphene channel is more symmetric compared to the four-terminal JJ presented in the main text. Here, we observe a nonreciprocal efficiency of $\eta^{\pm}_{12} \approx 7 \%$ which is smaller compared to the four-terminal device of the main text. This observation highlights the importance of mirror symmetry in determining the MTJJ’s nonreciprocity. To better understand the dependence of $\eta$ on the channel asymmetry, we perform numerical simulations based on resistively shunted junction (RSJ) model. Using the circuit-network model presented in the main text (inset of Fig.~\ref{P1.png}B), we characterize the broken spatial mirror symmetry as the ratio between the critical currents of JJ3 and JJ1/JJ2 (i.e., $x=I_{c3}/I_{c1}=I_{c3}/I_{c2}$). In this case, $x=1$ represents the perfect mirror symmetry. Our modeling shows that the efficiency increases with increasing $x$ and eventually plateaus at $\approx 30\%$ (Figure ~\ref{S7.jpg}). This maximum theoretical $\eta$ is comparable to our observed nonreciprocal efficiency of $\eta \approx 24\% $ in the three-terminal JJ.


\bibliography{References}

\newpage


\end{document}